# Computer Vision Approach to Study Surface Deformation of Materials

C. Zhu[a], H. Wang[b], K. Kaufmann[b], K.S. Vecchio[a,b*]

[a]Materials Science and Engineering Program, UC San Diego, La Jolla, CA 92093, USA

[b]Department of NanoEngineering, UC San Diego, La Jolla, CA 92093, USA

**Abstract**

Characterization of the deformation of materials across different length scales has continuously attracted enormous attention from the mechanics and materials communities. In this study, the possibility of utilizing a computer vision algorithm to extract deformation information of materials has been explored, which greatly expands the use of computer vision approaches to studying mechanics of materials and potentially opens new dialogues between the two communities. The computer vision algorithm is first developed and tested on computationally deformed images (% error <0.035%, L2-norm <2.5), before evaluating experimentally collected images on speckle painted samples before and after deformation. Moreover, a virtual experiment shows the feasibility of mapping surface strain of a sample based on its natural pattern with significantly improved accuracy compared to the digital image correlation result obtained from an open-source software Ncorr, which provides new opportunities in experimentation and computer algorithms to study deformation mechanics of materials. Validation experiments include evaluating the performance of strain mapping using the computer vision approach in the uniaxial tensile test and three-point bending test, compared with extensometer reading and digital image correlation respectively.

*Corresponding author: kvecchio@eng.ucsd.edu, (858) 822-7922, (858) 534-9553

**Nomenclature:**

$I_m(x, y)$: intensity of moving image at point $(x, y)$

$I_f(x, y)$: intensity of fixed image at point $(x, y)$

$\vec{u}$: displacement field to remap moving image towards fixed image

T: transformation between fixed image and moving image

$u_x$: displacement in X

$u_y$: displacement in Y

F: deformation gradient tensor

E: Green-Lagrange strain tensor

$E_{expected}$: expected Green-Lagrange strain tensor

A: Eulerian-Almansi strain tensor

$A_{expected}$: expected Eulerian-Almansi strain tensor

D: affine transformation matrix

$e_x$: engineering strain in x (image scale factor in X)

$e_y$: engineering strain in y (image scale factor in Y)

P: pyramid level

# 1. Introduction

The state-of-the-art in extracting feature variation in images is divided into two communities: the mechanics community and the computer vision community. Although these two communities share almost the same motivations, there has rarely been any dialogue between the two [1]. The mechanics community seeks highly robust and accurate methodologies to map strains of artificial patterns painted on the objects of interest for deformation studies, whereas the computer vision community endeavors to implement a fast and simple feature- or intensity-based algorithm on an objects' natural surface pattern or texture.

In experimental mechanics, understanding how materials deform, and where they are most susceptible to failure is crucial to promote safety and potentially guide material design to alleviate catastrophic loss. Characterization methods at different length scales that can map strain in-situ and probe residual deformation post-mortem are therefore very important to provide researchers with location-specific measurable quantities to study failure mechanisms.

One of the most commonly used in-situ strain mapping techniques within the mechanics community is the digital image correlation (DIC) technique. It is a well-established non-interferometric technique, first developed by Sutton *et al*. in the 1980s, which probes the full-field surface displacement and strains in-situ with sub-pixel spatial resolution and high accuracy [2–7]. High-spatial resolution DIC can be carried out in a scanning electron microscope if the spatial and drift distortions are properly corrected [8]. A similar method called particle image velocity (PIV) also exists in the experimental fluid mechanics community, which is used to track velocities of traveling particles [9]. Recently, new global DIC techniques have been proposed as new alternatives to the traditional DIC [10–14]. However, the DIC or global DIC approach does not fully resolve the strain tensor, and it is insensitive to out-of-plane or in-plane movement. Alternative techniques such as stereo-DIC method has been developed to track out of plane motion i.e. full field strain tensor of the surface based on a calibrated two-camera [15] or one-camera [16] imaging systems. Another three-dimensional variant of the DIC technique is the digital volume correlation (DVC) technique based on X-ray tomography [17]. The limitation of DVC is that it relies on natural patterns found within samples to track the 3D displacement fields. It would be a difficult method for materials such as dense metal alloys that lack inherent internal patterns. The requirement of a non-uniform surface feature is also a recognized limitation for conventional DIC. Alternative diffraction-based techniques such as X-ray [18], neutron diffraction [19], or electron backscatter diffraction [20] have been established to probe the full strain (as well as rotation) tensor at microscale or even nanoscale through diffraction of a small volume of crystalline material. However, these approaches require quite complex instrumentation and analysis techniques to extract the strain data.

In the computer vision community, feature based image registration such as the Lucas-Kanade's optical flow algorithm has been widely used since 1981 [21]. Many other applications have emerged over the years from the image registration community to deal with tracking [22], face coding [23], medical image registration [24], parametric and layered motion estimation [25], mosaic construction [26], etc. One of these many applications involves study of the deformation of objects by applying the *demons* algorithm on objects natural patterns [27–29]. The original *demons* algorithm was first developed by Thirion as an analogy of Maxwell's demon. In the *demons* algorithm, the use of contour is very close to an early 'anticipating snakes' algorithm developed by Ronfard for image segmentation [30]. The *demons* algorithm has later been widely implemented for many radiotherapy applications [31–35]. The authors of this paper have also recently used *demons* algorithm for pattern remapping in a HR-EBSD study to extract residual stress and strain information from an additively manufactured Inconel sample [36]. Several other groups have dedicated their efforts in establishing more in-depth theoretical framework for the *demons* algorithm [37–40]. In the methodology section, the *demons* algorithm is covered in more details.

Registration can also be made more efficient for small deformation by considering gradient information from the moving image to allow diffusion of the fixed image into the moving image. A new 'active' force, proposed based on Newtons' third law of motion, that 'diffuses' the fixed image to match the moving image is added by Rogelj and Lovacic [41] under the assumption of the bi-directional diffusional process. It has been shown that consistency and registration accuracy can be improved when both 'passive' and 'active' forces are treated in the same manner. Nevertheless, the *demons* algorithm is still limited to measuring small deformation. For example, the diffusing model-based *demons* algorithm would have a initialization problem to correctly define the 'inside' and 'outside' regions when the objects do not overlap i.e. large deformation. It means that the diffusing model-based *demons* method lacks sub-pixel sensitivity compared to digital image correlation.

To overcome the limitation of *demons* algorithm being only applicable for small deformation, a pyramidal (coarse-to-fine) multi-resolution approach is proposed by Kostelec *et al.* [42]. This approach utilizes low-resolution images derived from the original fixed and moving images to start the diffusional registration process. The displacement field derived from low-resolution images are then passed on to the next higher resolution level as a starting solution. The entire diffusion process includes separate iterative *demons* registration at each pyramid level, and proceeds from the lowest image resolution level to the highest image resolution level. Since the computation speed is faster for lower resolution images, the multi-resolution approach significantly improves calculation speed as well as convergence. Moreover, the *Spectral-Log Demons* algorithm has also been developed by Lombaert *et al.* to overcome the limitation of traditional *demons* algorithms to small deformation by capturing the large deformation through Spectral Forces [43,44]. Another interesting study dealing with large deformation by Zhao *et al*. uses a two-layer deep adaptive registration framework to separately classify the rotation parameter through convolutional neural networks and then identify scale and translation [45].

The intention of this study is not to create a novel image registration algorithm, but to implement an existing multi-resolution *demons* image registration algorithm and evaluate the performance of the method to resolve large deformation on both computationally deformed images and experimental data. The accuracy of the program is first tested in virtual deformation experiments with artificially patterned images deformed using affine transformations. From a series of parametric studies, the optimal parameters for image registration of heavily deformed images are determined based on the convergence of measured strain to expected strain in the virtual experiments. Then, the program is applied to real deformation experiments performed in the lab, including tensile deformation of pure iron (homogeneous deformation) and a three-point bending test (heterogeneous deformation). The measured axial strain in the tensile test is compared with extensometer results and the expected Green-Lagrange strain, whereas the measured heterogeneous strain in the bending test is compared with strain measured using DIC software.

## 2. Methodology

In the *demons* registration algorithm used in this paper, the image-to-image matching is effectively through a diffusing model, in which object boundaries in one of the images (fixed image) act as semi-permeable membranes (static contours) and the other image (moving image) contains a deformable grid that diffuses through these boundaries by the action of effectors (*demons)* along the membranes. The orientation of the demons force are defined along the static iso-contours ($\nabla I_f(x_0, y_0) \neq 0, I = I_f(x_0, y_0)$) on the fixed image from inside ($I_f(x, y) < I$) to outside ($I_f(x, y) > I$). The displacement field on the moving image is comparable to the *demons* force in that if the $I_m(x,y) > I$, the orientation of displacement is along $\nabla I_f(x_0, y_0)$, and when $I_m(x,y) < I$, the orientation of displacement is along $-\nabla I_f(x_0, y_0)$.

This method of computing the displacement field, or the 'passive' force to register the moving image (undeformed) to the fixed image (deformed), is the optical flow method. The optical flow method is originally used to determine small deformations in time-resolved images [10,46,47]. For a given point $(x, y)$, the intensity of the moving image $I_m(x, y)$ is hypothesized to be constant with time. Therefore, the difference between image intensities of moving and fixed images at the same point $(x, y)$ can be related to the displacement (from the point $(x, y)$ in the moving image to the point with same image intensity in the fixed image) and the intensity gradient in the fixed image [27].

$$\vec{u} \cdot \nabla I_f(x, y) = I_m(x, y) - I_f(x, y). \tag{1}$$

Eq.1 can be further modified to consider $\vec{u}$ to be the shortest spatial displacement that brings $I_m(x, y)$ of its hypersurface into hyper-plane coming through $I_f(x, y)$ [27].

$$\vec{u} = \frac{[I_m(x, y) - I_f(x, y)] \nabla I_f(x, y)}{\nabla I_f(x, y)^2}. \tag{2}$$

The displacement obtained from the above equation is clearly unstable when $\nabla I_f(x, y)$ approaches small values, which should ideally be close to zero. Hence, additional modification to Eq.2 is introduced that guarantees the behavior of displacement field at small $\nabla I_f(x, y)$.

$$\vec{u} = \frac{[I_m(x, y) - I_f(x, y)] \nabla I_f(x, y)}{\nabla I_f(x, y)^2 + [I_m(x, y) - I_f(x, y)]^2}. \tag{3}$$

where $\vec{u}$ is the displacement field, $I_m(x, y)$ is the intensity of point in the moving image at a position $(x, y)$, $I_f(x, y)$ is the intensity of point in the fixed image at the same spatial position $(x, y)$, and $\nabla I_f(x, y)$ is the intensity gradient i.e. 'internal force' of the fixed image at a position $(x, y)$. In some studies regarding the implementation of *demons* algorithm, an objective function is used as a registration metric (energy function), which is then optimized [34,38]. The registration metric (mean squared error) is then optimized to obtain the best transform T i.e. a convergence algorithm. The minimization of registration metric often involves another regularization term to avoid an ill-posed problem with unstable and non-smooth solutions [34]. In general, the minimization of this mathematical framework is computationally intensive. However, the *demons* algorithm by Thirion is a more efficient registration scheme, which does not necessarily guarantee convergence [27].

Mathematically, *demons* registration approach by Thirion aims to iteratively determine a final transform T ∈ $\mathcal{T}$ ($\mathcal{T}$ is a set of allowed displacement fields) between the space $\mathcal{M}$ of the moving image and the space $\mathcal{F}$ of the fixed image. The final transform $T$ (with an initial transform $T_0$ as the identity), which maps pixels in the moving image towards fixed image, is calculated iteratively (number of iterations=i) to allow convergence of the $I_m$ towards $I_f$. For instance, the transformed version of $I_m$ i.e. $T_i(I_m)$ is determined by the 'external' forces between $T_{i-1}(I_m)$ and $I_f$ as well as 'internal' forces that constrain the points in the moving image using Eq.3 [27].

$$T_i(I_m) = T_{i-1}(I_m) + \vec{u}_i(I_m). \tag{4}$$

The limitation is that if the initial fixed and moving images are drastically different (large deformation). Error in the displacement would gradually accumulate during the iterative registration process. To reduce the error in the displacement, the multiresolution *demons* framework first obtains accurate estimate of the large deformation from a pair of scaled-down fixed/moving images, which is then used as initial transform for the next pyramid level of scaled images. Details of this mutiresolution *demons* framework are discussed in the next section.

**2.1 Implementation of *Demons* Algorithm for Strain Mapping**

In this study, the *demons* algorithm is implemented in Matlab™; the workflow of the program is summarized in Fig. 1. The input consists of two images: one fixed image (deformed state) and one moving image (undeformed state). Before feeding the images into the *demons* solver, a histogram matching filter and anisotropic diffusion filter [48] are employed to balance the illumination and remove noise. For appropriately illuminated (diffuse light source) footage collected from experiments, the histogram matching filter is not necessarily required. Compared to a Gaussian filter, an anisotropic diffusion filter is more effective at preserving important image features such as edges, lines, etc., while effectively reducing image noises [48]. The two images are then down-scaled from their original resolution (at pyramid level $P_{n=0}$) according to the specified number of pyramid levels. Each added level contains a down-scaled image from the lower pyramid level by a factor of two along each coordinate direction. The image registration process begins from the highest pyramid level, i.e., lowest image resolution, and runs through an iterative process illustrated in Fig.1. The solution for the displacement field is then passed onto the lower pyramid level as an initial solution, which then significantly improves the accuracy of diffusional registration for large deformation. The final displacement solution is eventually generated from the lowest pyramid level containing images with the original resolution.

From the displacement field obtained through image registration in undeformed coordinates, it is then possible to obtain the Green-Lagrange strain tensor, a rotation independent tensor that describes the deformation of material points with respect to the undeformed configuration i.e. Lagrangian description. Since direct differentiation of the displacement field is sensitive to noise, a Savitzky-Golay filter [49,50] is first employed to smooth the displacement field. The SG-filter is a type of low-pass filter that works by least-square fitting 2D polynomial to a local subregion (defined by its odd number sized window). The displacement fields that need to be smoothed in this case contains two parts second-order polynomial): $u_x(X,Y)$ and $u_y(X,Y)$.

$$u_x(X,Y) = a_0 + a_1 X + a_2 X^2 + a_3 Y + a_4 XY + a_5 Y^2. \tag{5}$$

$$u_y(X,Y) = b_0 + b_1 X + b_2 X^2 + b_3 Y + b_4 XY + b_5 Y^2. \tag{6}$$

For a windows size of (2w+1) by (2w+1) (odd number, center=0,0), the above expression can be written in terms of matrix form Ax=b, where x is the column vector containing all the fitting coefficients:

$$\begin{pmatrix} 1 & -w & w^2 & -w & w^2 & w^2 \\ 1 & -w+1 & (w-1)^2 & -w & w(w-1) & w^2 \\ \vdots & \vdots & \vdots & \vdots & \vdots & \vdots \\ 1 & w-1 & (w-1)^2 & w & w(w-1) & w^2 \\ 1 & w & w^2 & w & w^2 & w^2 \end{pmatrix} \begin{pmatrix} a_0 \\ a_1 \\ a_2 \\ a_3 \\ a_4 \\ a_5 \end{pmatrix} = \begin{pmatrix} u_x(-w,-w) \\ u_x(-w+1,-w) \\ \vdots \\ u_x(w-1,w) \\ u_x(w,w) \end{pmatrix}. \tag{7}$$

The coefficient vector x can be obtained through the pseudo-inverse matrix $(A^T A)^{-1} A^T$: $x=(A^T A)^{-1} A^T b$. The smoothed value for $u_x(0,0)$ at the center of the window is $a_0$, and $b_0$ in the case of $u_y(0,0)$. By moving along the polynomial fitting window, the displacement field can be smoothed for $u_x(X,Y)$ and $u_y(X,Y)$ respectively.

The displacement gradient terms ($\frac{\partial u_x}{\partial X}, \frac{\partial u_x}{\partial Y}, \frac{\partial u_y}{\partial X}, \frac{\partial u_y}{\partial Y}$) are then computed locally using the central difference algorithm to obtain the corresponding deformation gradient tensor F (given by Eqn. 2) of every point within the selected area in the undeformed coordinates (deformed material point vectors: x, y; undeformed material point vectors: X and Y). For an experiment containing larger amounts of noise, the local subset plane fitting method is implemented instead of the central difference method [49].

$$F = \begin{pmatrix} \frac{\partial x}{\partial X} & \frac{\partial x}{\partial Y} \\ \frac{\partial y}{\partial X} & \frac{\partial y}{\partial Y} \end{pmatrix} = \begin{pmatrix} \frac{\partial (X+u_x)}{\partial X} & \frac{\partial (X+u_x)}{\partial Y} \\ \frac{\partial (Y+u_y)}{\partial X} & \frac{\partial (Y+u_y)}{\partial Y} \end{pmatrix} = \begin{pmatrix} 1+\frac{\partial u_x}{\partial X} & \frac{\partial u_x}{\partial Y} \\ \frac{\partial u_y}{\partial X} & 1+\frac{\partial u_y}{\partial Y} \end{pmatrix}. \tag{8}$$

For every point in the selected area, the full 2D Green-Lagrange strain tensor, E, derived from the deformation gradient tensor, F, is then given by Eqn. 3, which inherently removes the contribution from rigid body rotation through the right Cauchy-Green deformation tensor ($F^T \cdot F$). T here represents the matrix transpose operation.

$$E = \frac{1}{2}(F^T \cdot F - I). \tag{9}$$

Here, it is important to distinguish the Eulerian-Almansi strain (A) from the Green-Lagrange strain (E). It is, like Green-Lagrange strain tensor, a type of strain tensor calculated with respect to deformed sample coordinates i.e. Eulerian description. It is therefore related to the deformation gradient tensor (of the Lagrangian description) via a slightly different relationship:

$$A = \frac{1}{2}(I - (F \cdot F^T)^{-1}). \tag{10}$$

**2.2 Virtual Experiments with Computationally Deformed Images**

In this study, a series of parametric studies and validation experiments using computationally deformed images are carried out, see Fig. 2. A test image (moving image) is computationally deformed uniaxially in tension (engineering strain in the x direction, $e_x$>0) to obtain the deformed fixed image using image warping. The image warping used here is an inverse (backward) mapping of the integer-coordinate pixels in the warped image towards the real-coordinate pixels in the input image to avoid round-off error in the forward mapping method. For a given forward geometric transformation D, the spatial coordinates in the undeformed (X, input image) can be calculated through the multiplication of an inverse matrix of D and the deformed coordinates (x, warped image).

$$X = D^{-1}x. \tag{11}$$

The spatial coordinates X corresponding to the integer coordinates in x is very likely not integer coordinates. Therefore, the intensity of pixels in the warped image can then be interpolated from neighboring pixels of the input image with bi-cubic interpolation method [51].

$$D = \begin{pmatrix} 1+e_x & 0 \\ 0 & 1+e_y \end{pmatrix} = \begin{pmatrix} 1+e_x & 0 \\ 0 & 1-0.3e_x \end{pmatrix}. \tag{12}$$

The amount of deformation is controlled by defining a type of affine transformation matrix in Eq. 12; in this case, the scaling matrix is used. For example, tensile deformation in the x-direction can be defined by assigning scale factors ($e_x$ and $e_y$) on the diagonal terms. This test image is assumed to be taken from a hypothetical linear elastic material, which demonstrates significant elasticity, such that the applied longitudinal tensile strain can be easily related to the engineering strain $e_y$ in the transverse contraction direction through the Poisson's ratio (taken as 0.3, for this example).

The expected in-plane Green-Lagrange strain tensor due to the computationally-applied tensile deformation can, therefore, be related to the applied affine transformation tensor:

$$E_{expected} = \frac{1}{2}(D^T D - I) = \begin{pmatrix} E_{xx} & E_{xy} \\ E_{yx} & E_{yy} \end{pmatrix}. \tag{13}$$

The expected Eulerian-Almansi strain can be similarly obtained through:

$$A_{expected} = \frac{1}{2}(I - (D \cdot D^T)^{-1}). \tag{14}$$

To evaluate the error between expected strain and measured strain, the percentage error is introduced. For instance, the percentage error for $E_{xx}$ is given by:

$$\% \, error = 100 \times \frac{measured \, E_{xx} - expected \, E_{xx}}{expected \, E_{xx}}. \tag{15}$$

The measured $E_{xx}$ is an average value of the measured $E_{xx}$ map, which also has a standard deviation (encloses 68.2% of the data) that informs the spread of measured strain data. The standard deviation is conveniently used as error bar of the average $E_{xx}$.

Another important error metric, L2-norm, is also calculated between the measured $E_{xx}$ map and expected $E_{xx}$ map (size of map W by H), which provides more information regarding the spatial correlation of the strain maps.

$$\text{L2} - \text{norm} = \sqrt{\sum_{y=1}^{H}\sum_{x=1}^{W}(measured\ E_{xx}(x,y) - expected\ E_{xx}(x,y))^2}. \tag{16}$$

**2.3 Validation Experiments with Tensile Test and Three-Point Bending Test**

Spray painting a black on white speckle pattern is done using a Paasche H series Airbrush and CREATEX airbrush paints. A uniform white layer is first painted on the polished sample before applying the random black speckle patterns. The black speckle size distribution in this study varies from 5 μm to 30 μm, which is sufficiently fine for the intended experiments.

A primary validation experiment was carried out to assess the performance of multi-resolution image registration on a uniformly deformed sample under tension. In this case, quasi-static tensile deformation of a dog-bone shaped pure iron sample, spray-painted with black and white speckles, has been recorded with a Supereyes microscope camera (Model: B011) at a resolution of 1600 by 1200 pixels. An axial extensometer is used to measure the tensile strain of the recorded area of interest as strain validation data. The tensile test is manually stopped at ~0.15 engineering strain before the axial strain reaches the limit of the extensometer.

Another validation experiment has been carried out to assess the performance of multi-resolution image registration on large-scale heterogeneous deformation. A pre-notched stainless steel Charpy impact bar is subject to three-point bending under quasi-static loading condition. The video of the deforming sample was recorded using a different camera setup, which consists of a NAVITAR's body tube lens mounted on an AMSCOPE's CMOS 5.1MP camera (model MU500).

**3. Results and Discussion**

## 3.1 Expected Strains for Computationally Deformed Images

A test image taken from a spray-painted sample is used for the following validation study for computationally deformed images. Based on the applied affine transformation, D, defined for computationally deformed images in tension ($e_x \in [0, 0.3]$), several expected measures of strains can be readily calculated as shown in Fig. 3. The engineering tensile strain $e_x$ can be computed from measuring the change in deformed image dimension relative to the undeformed image. In this study, the measured engineering tensile strain (red dashed line) is numerically equal to the applied scale factor (black solid line) in the affine transformation matrix, as shown in Fig. 3. In addition, the expected true strain value (blue dashed line) can be calculated using $e_t = \ln(e_x+1)$, which sits below the engineering strain. Furthermore, the expected Green-Lagrange strain (green dashed line) and the expected Eulerian-Almansi strain (brown dashed line) is obtained using Eq. 13. and Eq. 14, respectively, which differ significantly from the engineering strain. For the following parametric studies, the Green-Lagrange strain is calculated using the *demons* algorithm-based image registration code developed in this study and compared to the expected Green-Lagrange strain for percentage error/L2-norm analysis.

## 3.2 The Effect of Pyramid Level on Image Registration

Determining the appropriate number of pyramid levels used for image registration is crucial in obtaining a reliable displacement field. In this part of the parametric study, the most deformed image ($e_x = 0.3$) and the undeformed image ($e_x = 0$) are used to study the minimum number of pyramid levels required to correctly obtain the displacement field by varying the number of pyramid levels used from one to ten. Expected values for the Green-Lagrange tensor can be calculated using Eq. 13. For each of the registrations, the other necessary parameters such as number of iterations and SG-filter window size are kept constant throughout. At each pyramid level, 100 iterations are used, which is more than enough to converge to an accurate solution according to the next part of the parametric study. In addition, an SG-filter window size of 21 by 21 pixels is adopted. In all the parametric studies, the measured Green-Lagrange tensor values are average values of the different strain components within the map. Care must be taken near the edges of the image where the gradient of displacement field becomes erroneous.

As shown in Fig. 4, the minimum number of pyramid levels required to reach an error below 0.1% is eight for all the measured components of the Green-Lagrange strain tensor. The top two rows of images in Fig. 4 represent the difference in image intensity values between the registered undeformed image and deformed image at each different pyramid level. The image intensity difference is a simple subtraction of the two. The intensity difference plots indicate that the registration of the undeformed image towards deformed is quite poor at lower pyramidal levels between (a) to (g) i.e. pyramid level one to eight. It clearly shows improved registration when the number of pyramid levels is equal to or greater than eight. At lower pyramid levels, the assumption in the optical flow equation that the 'forces' are derived from close features between fixed and moving images leads to erroneous estimation of the displacement field. By increasing the number of pyramid levels, the features become blurred and merged into 'closer' features that are more tractable with the *demons* algorithm. Additionally, the error bar of the plots indicates the standard deviation of the measured strain components. It is observed that the error bars become unnoticeable as soon as the pyramid level reaches eight. In addition, the L2-norm drops significantly at and beyond pyramid level eight, meaning the measured strain maps are well correlated with expected strain maps.

### 3.3 The Effect of Number of Iterations at Each Level on Image Registration

The diffusional image registration process is an iterative approach, which eventually allows the matching of the features. Since no convergence criteria is implemented in this study, the minimum number of iterations required to achieve accurate matching of the images is therefore very important to the determination of the strain tensor. In this section of the parametric study, the number of iterations is taken from 1 to 90 with an interval of 10 to match the most deformed image ($e_x = 0.3$) and the undeformed image ($e_x = 0$). The SG-filter window size is 21 by 21 pixels, and the number of pyramid levels is eight. It has been shown in Fig. 5. that the components of measured Green-Lagrange strain maps correlate nicely with the expected strain maps, when the number of iterations used at each pyramid level exceeds 40. The error bars also clearly show that the strain maps of the three tensor components immediately become uniform at 40 iterations. Increasing the number of iterations beyond 40 slightly improves the convergence to expected values, but at the expense of more computational cost. GPU-based registration is also possible to implement by creating the corresponding GPU arrays of the images in the Matlab program and then carry out the registration process, which could significantly reduce run time.

**3.3 The Effect of Savitzky-Golay Filter's Window Size on Image Registration**

The Savitzky-Golay filter is applied to smooth the displacement field before strain calculation. It should be exercised with care for heterogonous deformation so that it does not remove details representing the deformation. Ideally, the filter size should be smaller than the smallest deformation feature for heterogeneously deformed samples. In this part of the parameter study, the square SG filter's window size is varied from 3 by 3, 5 by 5, 7 by 7, 9 by 9, 13 by 13, 15 by 15, 21 by 21, 27 by 27, 35 by 35 and 41 by 41. The pyramid level is fixed at eight and the number of iterations used is 40. The most deformed image ($e_x = 0.3$) and the undeformed image ($e_x = 0$) are fed into the image registration program as input. As shown in Fig. 6, the accuracy of strain calculation clearly does not depend on the SG-filter's window size for the virtual tensile experiment in this study, whereas the precision of the strain field is slightly improved as shown by the error bars. An accuracy of less than 0.1% can be achieved with a window size of greater than 13 by 13. The percentage error of the average measured strain all fall below 1% and steadily converges to 0.02% at increasing window size. Moreover, the L2-norm indicates improved correlation between the expected strain maps and measured strain maps as windows size increases, presumably due to the smoothing effect that homogenizes the measured strain maps.

**3.4 Virtual Tensile Test Experiment Part I: Artificial Pattern**

A series of computationally deformed images, with spray-painted pattern, in tension are produced using an image warping algorithm with a predefined affine transformation matrix. The level of applied engineering strain on the image ranges from 0 to 0.3. Based on the previous parametric studies, the optimal parameters used for this virtual tensile experiment are determined to be: pyramid levels = 8, SG filter's window size = 21 by 21, number of iterations = 40.

The image registration approach is then applied to map the Green-Lagrange strain tensor at each applied uniform strain level. The measured average strain values for the different components of the Green-Lagrange tensor are then compared with expected results calculated using Eq.13 for percentage error analysis as shown in Fig. 7. Percentage error analysis indicates that error falls below 0.035% for all applied strain levels. The highest error is measured at the smallest applied strain level, which suggests that the technique is prone to error at extremely small. A similar study has been carried out at strain levels below 0.01 to reveal the limitation of the current approach (see Appendix Fig. A1). The percentage error associated with Green-Lagrange strain tensor for extremely small applied deformation can reach as high as 8% when the actual change in the pixel is close to one pixel. Overall, the correlation between the measured and expected strain maps is very well from the L2-norm analysis in both Fig.7 (c) and Fig.A1 (c). However, the strain error at small applied strains is not very explicitly shown because the absolute difference between the expected strain and measured strain is small at small applied strains without any normalization.

The percentage error associated with measuring the Green-Lagrange strain tensor using the image registration approach has also been assessed by imposing an additional 5° clockwise rotation to the deformed image, and compared with results obtained from Ncorr software (v1.2) developed by Blaber *et al*. [52]. Parameters used for Ncorr can be found in Appendix (Table 1). In this software, the normalized cross-correlation is used as an initial guess prior to the application of the nonlinear optimization algorithm, i.e., Inverse-Compositional Gauss-Newton (IC-GN) method to calculate displacement gradients [6]. Although the Ncorr software implements a more sophisticated approach to calculate full-field displacement/strain, it is referred to here as 'cross-correlation' in the figures for brevity. It has been shown in Fig. 8 that both methods accurately extract the strain tensor from the undeformed reference frame. The errors of all strain components are well below 1%, except for the $E_{yy}$ at the smallest applied strain. Overall, the Ncorr software seems to be slightly more accurate compared with image registration approach due to the advantage of IC-GN in terms of image alignment and efficient performance in image rotation experiment [53].

**3.5 Virtual Tensile Test Experiment Part II: Natural Pattern**

Another case study was performed on a series of computationally deformed images with natural pattern, e.g. grain boundary microstructure and back-scattered electrons' mass contrast using the same affine transformation matrices from section 3.4. This image, shown in Figure 9, is taken from a well-polished metallic-intermetallic laminate (MIL) composite surface using a scanning electron microscope (SEM). Similarly, the image registration parameters used for this virtual tensile experiment on natural patterns are pyramid levels = 8, SG filter's window size = 21 by 21, number of iterations = 40. To demonstrate the advantage of using the image registration approach over DIC, the Ncorr software (v1.2) is used as a comparison. A table containing the parameters used in the DIC software can be found in the Appendix (Table 1).

As shown in Fig. 9, the schematic at the top includes several (not all) deformed images using affine transformation. The percentage errors associated with the calculated Green-Lagrange strain tensor derived from image registration and cross-correlation (Ncorr) have been plotted in three separate subplots for the three components in log scale for the y-axis. It has been shown that both methods are capable of accurately measuring the applied strain tensor, whereas the image registration clearly demonstrates significant improvement in percentage errors over the Ncorr software by an order of magnitude. This virtual experimental suggests that the application of image registration approach can more accurately measure the surface strain of samples with a properly exposed natural pattern. It is also worth noting that the performance of *demons* is improved on the natural pattern compared to speckle pattern. This is because *demons* is established based on smooth image intensity gradients, whereas speckle patterns contains sharp changes that will influence the accuracy of the estimated displacement field.

**3.6 Experimental Validation Part I: Tensile Deformation of Pure Iron**

The method used to calculate the surface axial tensile strain is slightly modified from the previous parametric study in Section 3.4. Twenty-nine images are first extracted at an equal time interval from the video footage. The significant shift in the field of view in recording the video does not allow for accurate image registration between the first undeformed image and subsequent deformed images. Hence, image registration is conducted on pairs of neighboring images instead of using the first image as the reference image. Since the tensile deformation is stopped before the sample starts to neck, the deformation of the recorded part of the sample is uniform. The cumulative tensile strain can be calculated and compared with the extensometer result and the expected Green-Lagrange strain. The parameters used to carry out the image registration are: pyramid levels = 8, SG filter's window size = 21 by 21, number of iterations = 40. Since the deformation is uniform throughout, a local subset plane fitting method is adopted here to calculate the displacement gradient terms with a fitting window size of 5 by 5 [49].

It is shown that the image registration matches well with the extensometer result over the entire range in Fig. 10. However, the image registration only matches with the expected Green-Lagrange strain until ~0.05. The discrepancy between the measured Green-Lagrange strain from image registration and expected Green-Lagrange strain is quite significant above 0.05 strain. The reason for this discrepancy is that the sample surface is a free surface, which allows out-of-plane or in-plane movement or rotation of grains in a sample. The speckles painted on a grain with significant out-of-plane or in-plane motion will be registered to have lower than expected strains, leading to the strain being underestimated in comparison to the expected Green-Lagrange strain. Existing methods such as holographic interferometry is one possible method to extract the out-of-plane or in-plane displacements, but it is beyond the scope of this work [54].

**3.7 Experimental Validation Part II: Three-Point Bending**

In this experiment, a comparison of strain fields mapped using DIC and image registration on a pre-notched stainless-steel Charpy impact bar under three-point bend testing is presented. The DIC software (Ncorr v1.2) used in this part of the study is developed by Blaber *et al*. [52]. A table containing the parameters used in the DIC software can be found in the Appendix (Table 1). Similarly, the parameters used for image registration are: pyramid level = 8, SG filter's window size = 25 by 25, number of iterations = 70.

In Fig. 11, diffusional image matching through image registration shows that most of the image areas can be successfully registered except for a few edge regions. In this figure, the undeformed image is compared with the registered deformed image, since the Green-Lagrange strain in fact demands displacement gradients to be calculated in the undeformed coordinates. The histograms showing the difference in the intensity values (residual intensity values) between registered and unregistered pairs of images, which indicate a significant proportion of the image intensities (in grayscale) match up very well. The orange histogram contains both the un-matched edge regions, as well as minor intensity mismatch within the registered 'black' region, i.e., the registered region is not entirely black.

From the registered images, the displacement field that deforms the undeformed image to match the deformed image can be extracted and plotted as a 2 by 1 vector in Fig. 12. This displacement vector can be compared with the displacement field calculated from cross-correlation method with matching scale bars in units of pixels. The area of interest selected in the two methods are slightly different, but they cover almost the same region underneath the notch. It can be seen from the displacement in the x-direction that the material surrounding the notch has been pushed away from the notch, meaning that the notch has been opened up. However, the material closer to the loading pin gets compressed towards the pin, which corresponds to the compressive strain $E_{xx}$ at the bottom. Since the loading is applied axially at the center, opposite to the side with the notch, material underneath the notch gets pushed in the direction of the applied load (up in the image coordinates, but down in the sample coordinates). Material away from the notch also moves along to the direction of loading due to the fixed points by the side pins. This large-scale movement of material near the notch creates the large shear strain distribution $E_{xy}$ across the thickness of the sample as shown in Fig. 13. The strain distributions for $E_{xx}$ and $E_{yy}$ in Fig. 13 share similar shapes because the tensile strain in the x-direction must be accompanied by a corresponding compressive strain in y to preserve the volume of material. Based on the values of strain, the largest tensile strain, $E_{xx}$, is likely to result in the final failure of the Charpy bar.

The probability distributions for the different components of strains obtained from two different methods are directly compared. As shown in Fig. 13, the distributions for $E_{xx}$ and $E_{yy}$ are very similar to each other. The slight difference close to zero strain is likely due to the slightly larger region of interest chosen in the cross-correlation based method. Another clear discrepancy is observed in the shear strain, $E_{xy}$, distribution, where the cross-correlation method resolves more area in the sheared region, whereas the sheared region obtained from the image registration approach in Fig. 13 clearly contains a more discrete shear strain distribution.

### 3.8 Applications and Limitations

Application of multi-resolution *demons* algorithm in mapping surface strain of homogeneously/heterogeneously deformed material using simple optics setup is demonstrated to be quite feasible. With simple post-processing techniques such as histogram intensity adjustment and anisotropic diffusion filtering, it is possible to deal with poor illumination conditions and noise in the imaging system. Since the *demons* algorithm is originally used in analyzing biomedical images, unsurprisingly, this method can not only work with spray-painted samples, but also samples with a unique natural pattern, such as grain boundary microstructure features, mass contrast, etc. Therefore, it potentially allows mapping of strains of samples that are challenging to apply the artificial pattern on, for example, thin and soft sample, wet/reactive material, mechanically fragile sample, etc. Nevertheless, the current multiresolution *demons* approach apparently lacks accuracy at very small deformation (<0.01) when sub-pixel level sensitivity is desired. Moreover, further research is still required to enhance the robustness of this approach for experiments with a significant shift in the field of view and to implement an automated method to adapt the multiple parameters used for registration to the imaging condition, as well as the amount of deformation.

## 4. Conclusion

It has been demonstrated in this study that the computer vision approach, i.e., multi-resolution *demons* algorithm is effective in mapping surface strain with high accuracy on computationally deformed images painted with speckles used for DIC (% error <0.035%, L2-norm<2.5). A virtual experiment shows that it can also be used to map surface strain of samples based on its nature pattern, i.e. grain boundary microstructure and mass contrast, with an order of magnitude lower error in strain compared with DIC. In addition, experimental validation performed on experimentally collected video footage collected from spray-painted metallic samples deformed homogenously (tensile test) or heterogeneously (bending test) show strain values comparable to the extensometer reading and DIC based strain maps, respectively.

**Acknowledgment**

We would like to acknowledge Cheng Zhang, Tyler Harrington, and Wayne Neilson for assistance in the experimental setup. C. Zhu would like to acknowledge the Dissertation Year Fellowship provided by the home department. K. Kaufmann was supported by the Department of Defense (DoD) through the National Defense Science and Engineering Graduate Fellowship (NDSEG) Program. K. Kaufmann would also like to acknowledge the support of the ARCS Foundation.

**Appendix:**

Table 1: Cross-correlation parameters used in Ncorr software (v1.2)

|  | Section 3.4 | Section 3.5 | Section 3.7 |
|---|---|---|---|
| Image Resolution | 1113*644 | 854*603 (undeformed) | 2592*19444 |
| Analysis Type | regular | regular | regular |
| RG-DIC Radius | 25 | 25 | 25 |
| Strain Radius | 5 | 5 | 5 |
| Subset Spacing | 2 | 2 | 3 |
| Diffnorm Cutoff | 1e-06 | 1e-06 | 1e-06 |
| Iteration Cutoff | 50 | 50 | 50 |
| Step Analysis | Enabled (Seed Propagation) | Enabled (Seed Propagation) | Disabled |
| RG-DIC Subset Truncation | Disabled | Disabled | Disabled |
| Strain Subset Truncation | Disabled | Disabled | Disabled |

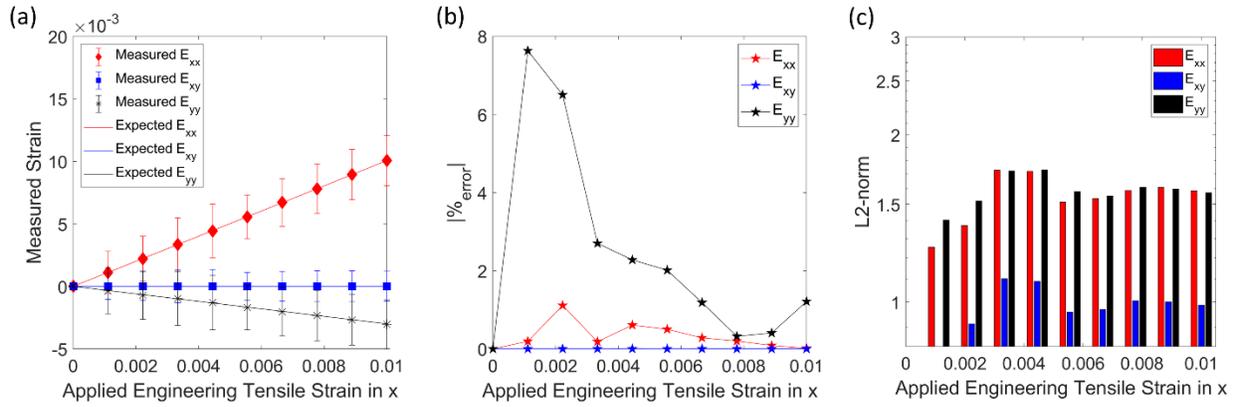

Fig. A1: (a) the measured average Green-Lagrange strain tensor components plotted against the expected strain tensor values, (b) the percentage error between the measured strain tensor components and expected strain tensor components, and (c) L2-norm between measured strain maps and expected strain maps as a function of applied engineering tensile strain (0-0.01) in the x-direction. (applied tensile strain in x is 0.3, number of iterations at each pyramid level is 40, number of pyramid levels is eight, SG-filter window size is 21 by 21 pixels).

**Figure captions:**

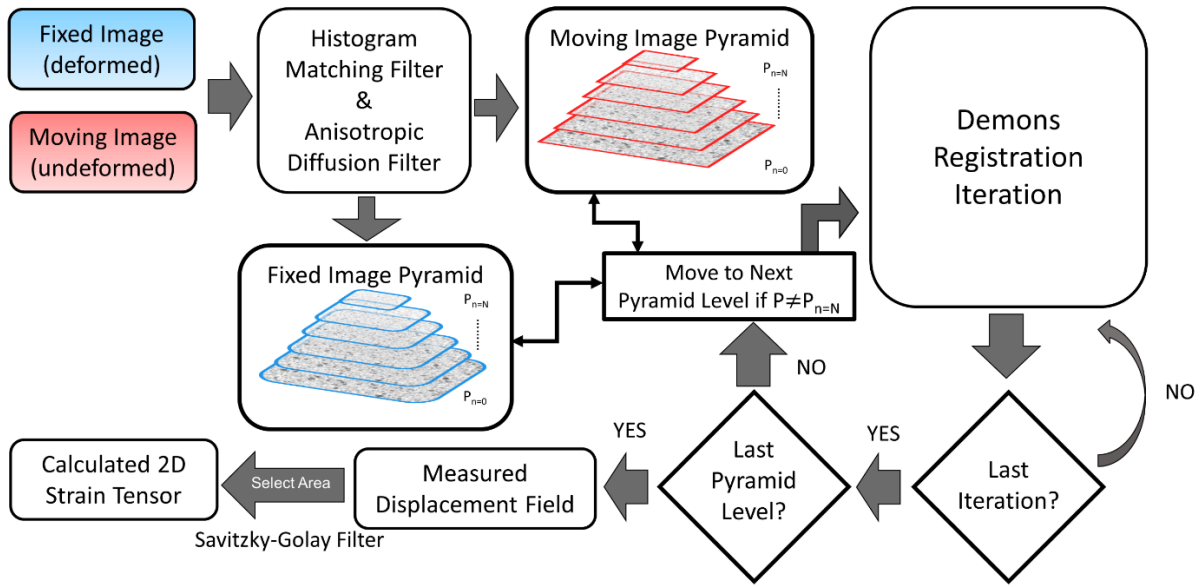

**Fig. 1** Flowchart of the multi-resolution image registration software developed based on the 'demons' algorithm.

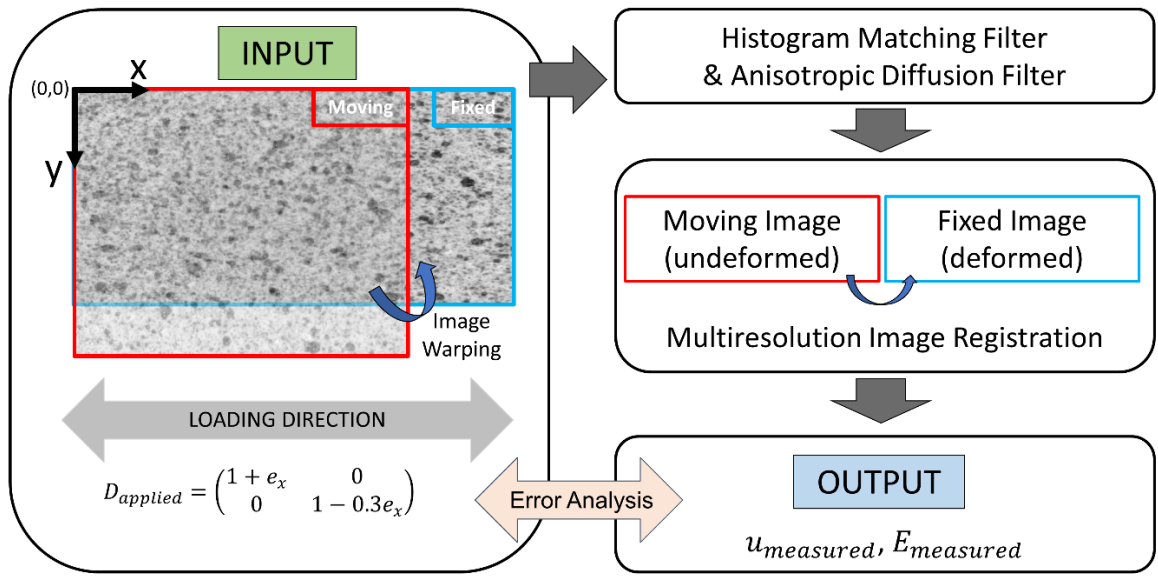

**Fig. 2** Flowchart of virtual deformation experiments with affine transformation and the validation of measured Green-Lagrange strain tensor with applied strain tensor.

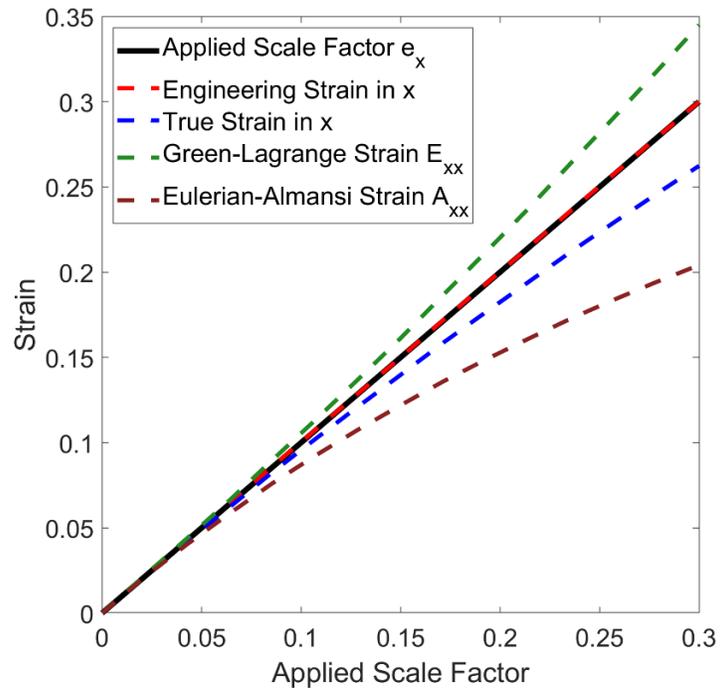

**Fig. 3** Different types of expected strain measures under the applied scale factor in affine transformation: engineering strain (red), true strain (blue), Eulerian-Almansi strain (brown) and Green-Lagrange strain (green).

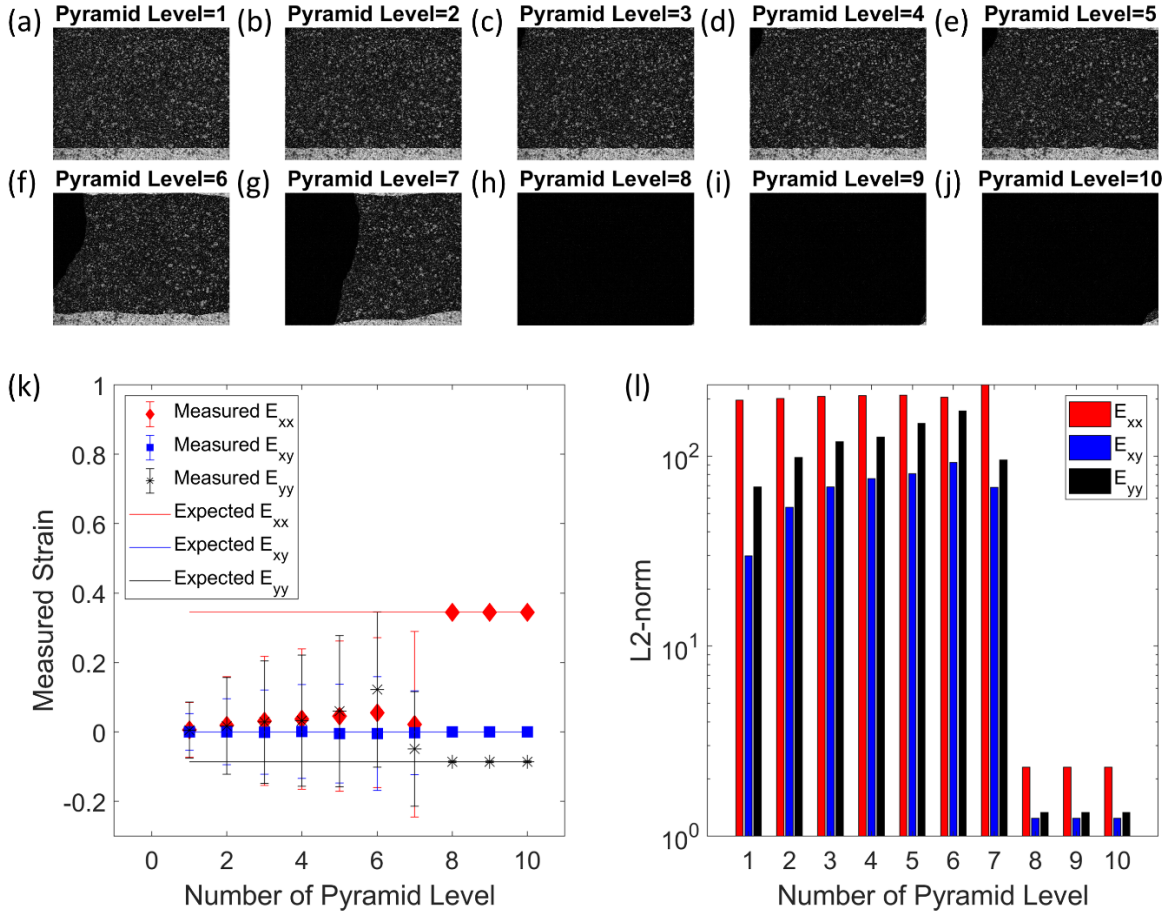

**Fig. 4** (a-j) Effect of using different number of pyramid levels on accuracy of image registration. Top two rows represent the intensity difference of undeformed image and registered deformed image at different number of pyramid levels, (k) the measured average Green-Lagrange strain tensor components plotted against the expected strain tensor values as a function of number of pyramid levels, and (l) the L2-norm between measured strain maps and expected strain maps as a function of number of pyramid levels. (SG-filter window size is 21 by 21 pixels, number of iterations at each pyramid level is 100).

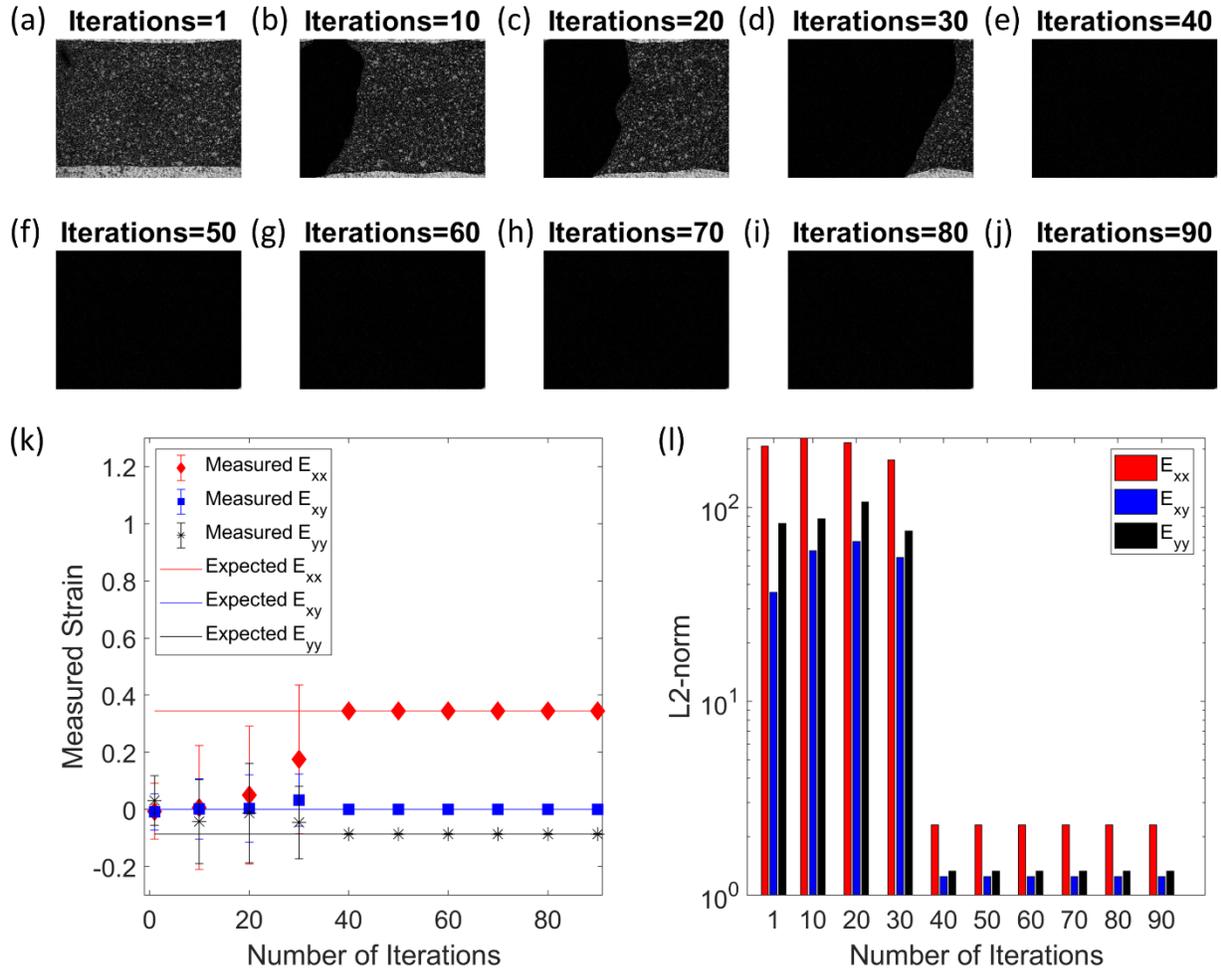

**Fig. 5** (a-j) Effect of using different number of iterations at each pyramid level on accuracy of image registration, (k) the measured average Green-Lagrange strain tensor components plotted against the expected strain tensor values as a function of number of iterations at each pyramid level, and (l) the L2-norm between measured strain maps and expected strain maps as a function of number of iterations for a given pyramid level. (SG-filter window size is 21 by 21 pixels, number of pyramid levels is eight).

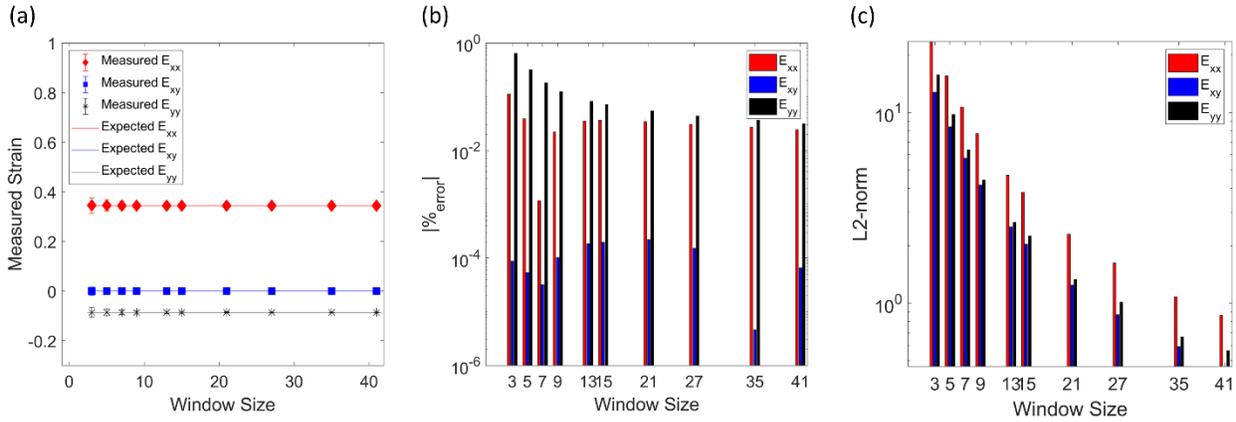

**Fig. 6** Effect of using different sizes of Savitzky-Golay filter windows on accuracy of image registration, (a) the measured average Green-Lagrange strain tensor components plotted against the expected strain tensor values, (b) the percentage error between measured strain tensor components and expected strain tensor components and (c) L2-norm between measured strain maps and expected strain maps as a function of SG filter window size. (number of iterations at each pyramid level is 40, number of pyramid level is eight).

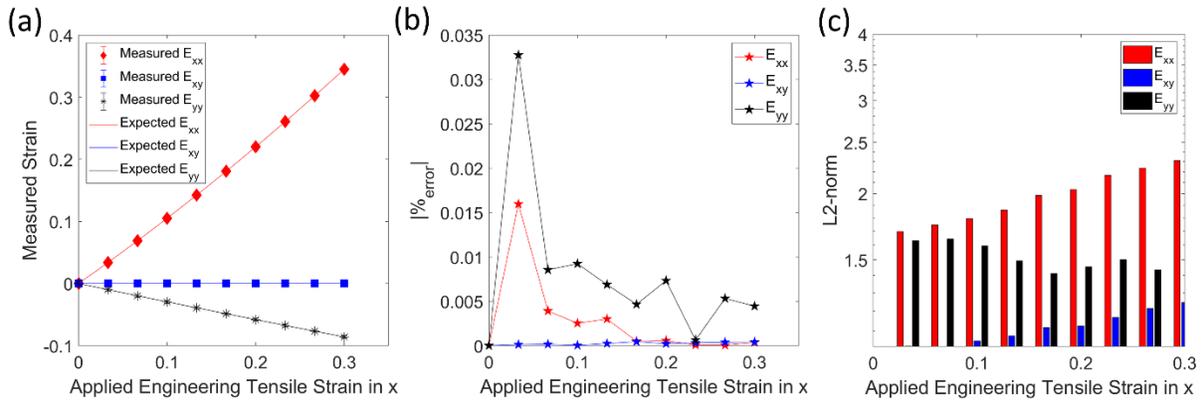

**Fig. 7** (a) the measured average Green-Lagrange strain tensor components plotted against the expected strain tensor values, (b) the percentage error between measured strain tensor components and expected strain tensor components and (c) the L2-norm between measured strain maps and expected strain maps as a function of applied engineering tensile strain in the x-direction (0-0.3),. (number of iterations at each pyramid level is 40, number of pyramid levels is eight, SG-filter window size is 21 by 21 pixels).

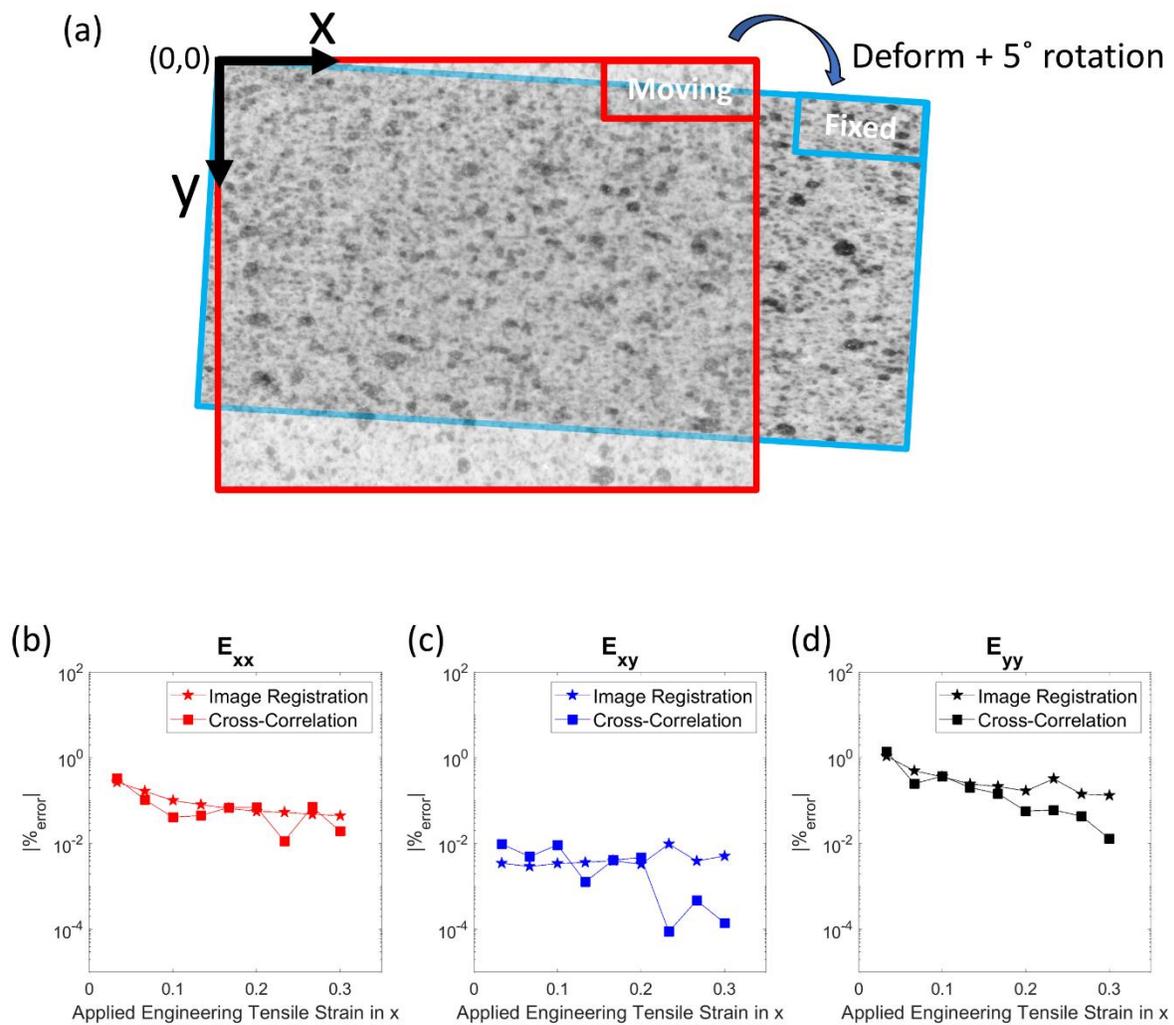

**Fig. 8** (a) schematic of the affine transformation (scaling matrix) of spray-painted images plus a 5˚ rotation; (b-d) percentage errors of average Green-Lagrange strain tensor components $E_{xx}$, $E_{xy}$, and $E_{yy}$ obtained from image registration and cross-correlation (Ncorr) methods.

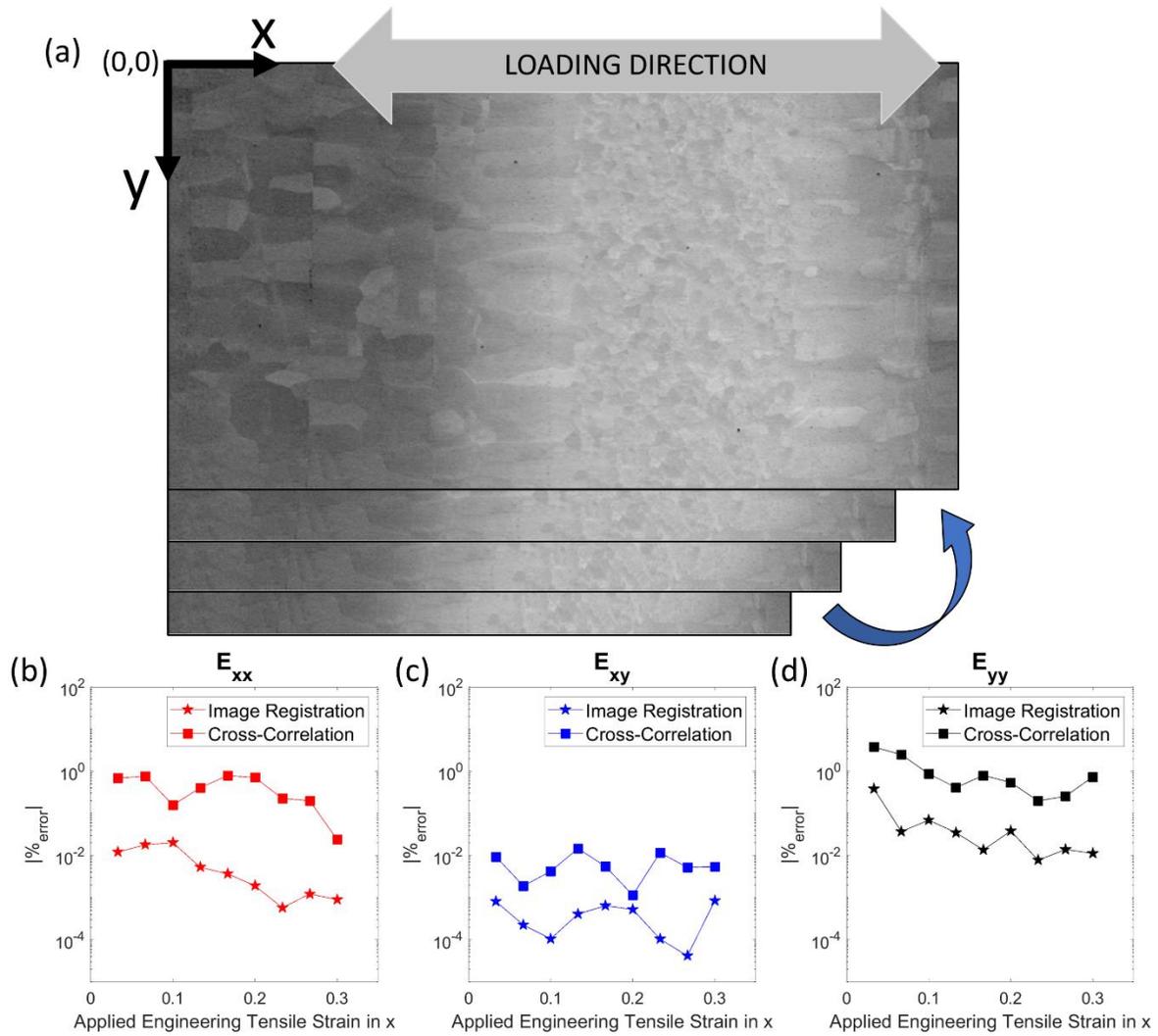

**Fig. 9** (a) schematic of the affine transformation (stretch) of an SEM taken from a metallic-intermetallic laminate component sample; (b-d) percentage errors of average Green-Lagrange strain tensor components $E_{xx}$, $E_{xy}$, and $E_{yy}$ obtained from image registration and cross-correlation (Ncorr) methods.

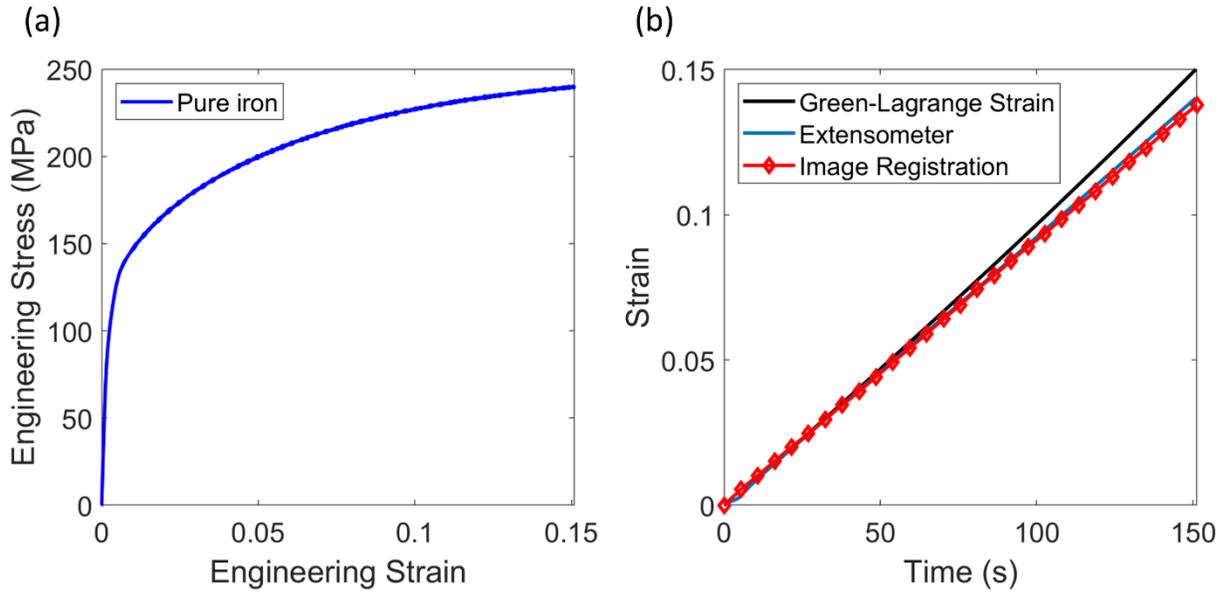

**Fig. 10** (a) the engineering stress-strain curve directly obtained from load frame, and (b) comparison between measured strain from both extensometer and image registration approach along with the expected Green-Lagrange axial strain.

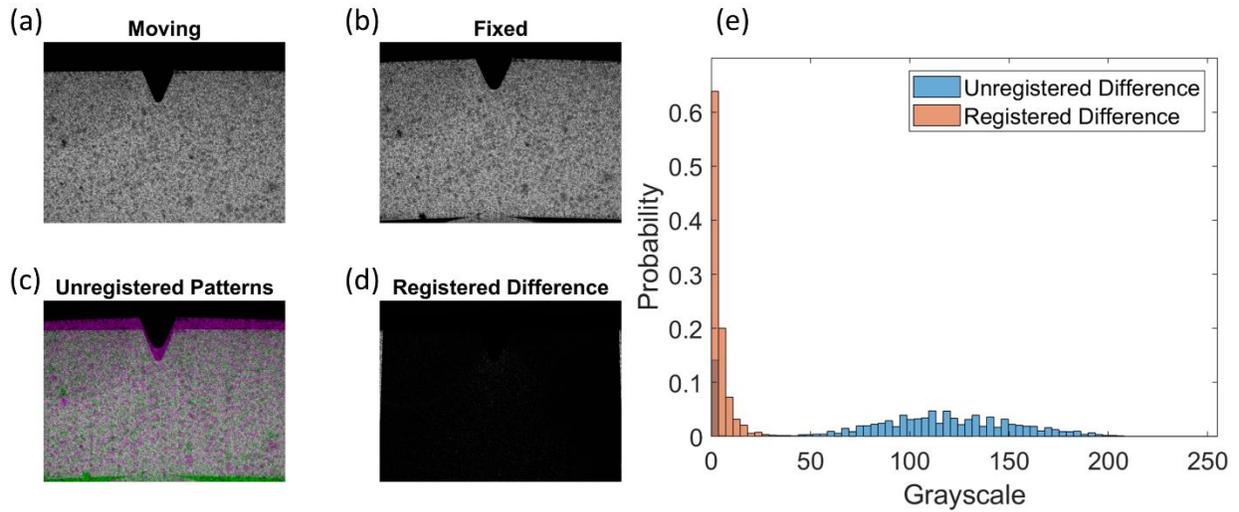

**Fig. 11** (a-d) the image intensity differences between unregistered deformed (fixed) and undeformed (moving) images as well as registered deformed image and undeformed image, and (e) histograms of intensity differences of the unregistered pair of images and registered pair of images.

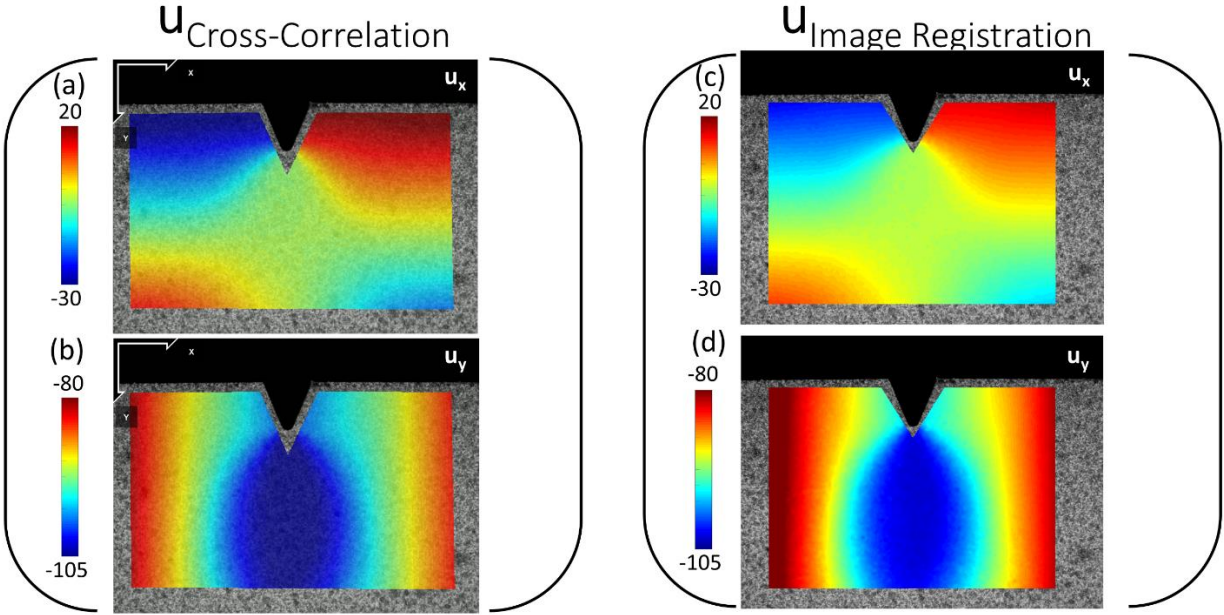

**Fig. 12** Displacement fields mapped using (a-b) cross-correlation-based DIC (Ncorr) and (c-d) multi-resolution image registration in a three-point bending test.

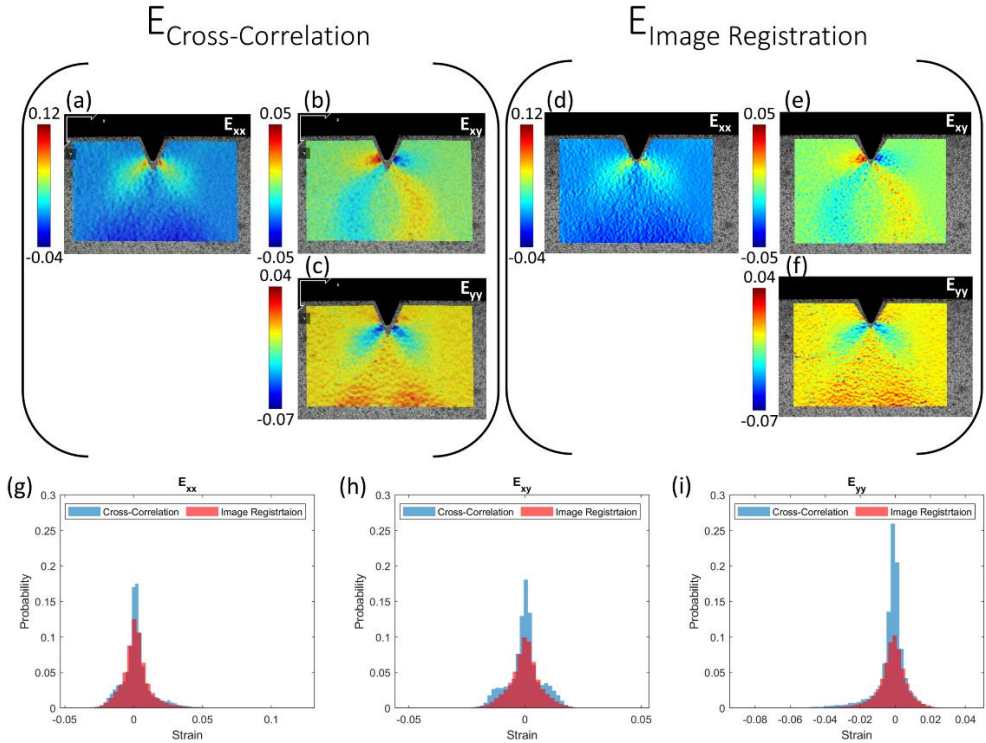

**Fig. 13** Green-Lagrange strain fields mapped using (a-c) cross-correlation based DIC (Ncorr) and (d-f) multi-resolution image registration in a three-point bending test, and (g-i) the comparison of histograms of strain fields mapped using cross-correlation based DIC (Ncorr) and multi-resolution image registration in a three-point bending test.

38